\documentclass[letterpaper]{article} % DO NOT CHANGE THIS
\usepackage{aaai2026}  % DO NOT CHANGE THIS
\usepackage{times}  % DO NOT CHANGE THIS
\usepackage{helvet}  % DO NOT CHANGE THIS
\usepackage{courier}  % DO NOT CHANGE THIS
\usepackage[hyphens]{url}  % DO NOT CHANGE THIS
\usepackage{graphicx} % DO NOT CHANGE THIS
\usepackage{subcaption}
\usepackage{natbib}  % DO NOT CHANGE THIS AND DO NOT ADD ANY OPTIONS TO IT
\usepackage{caption} % DO NOT CHANGE THIS AND DO NOT ADD ANY OPTIONS TO IT
\usepackage{algorithm}
\usepackage{algorithmic}
\usepackage{xcolor}
\usepackage[most]{tcolorbox}
\usepackage{siunitx}
\usepackage{amssymb}
\usepackage{pifont} 

\usepackage{newfloat}
\usepackage{listings}

\usepackage{bbding} 
\usepackage[capitalize,noabbrev]{cleveref}
\usepackage{booktabs}
\usepackage{multirow} 
\usepackage{amsfonts}

\usepackage{subcaption}

\definecolor{frame}{HTML}{7D9263}

\DeclareCaptionStyle{ruled}{labelfont=normalfont,labelsep=colon,strut=off} 
\floatstyle{ruled}
\newfloat{listing}{tb}{lst}{}
\floatname{listing}{Listing}

\title{Careful Queries, Credible Results: Teaching RAG Models Advanced Web Search Tools with Reinforcement Learning}

\author{
    Yuqin Dai\textsuperscript{\rm 1,5} \equalcontrib,
    Shuo Yang\textsuperscript{\rm 2}\equalcontrib,
    Guoqing Wang\textsuperscript{\rm 5}\equalcontrib,
    Yong Deng \textsuperscript{\rm 5},
    Zhanwei Zhang\textsuperscript{\rm 3,5},\\
    Jun Yin\textsuperscript{\rm 1}, 
    Pengyu Zeng \textsuperscript{\rm 1}, 
    Zhenzhe Ying \textsuperscript{\rm 5},
    Changhua Meng \textsuperscript{\rm 5},\\
    Can Yi \textsuperscript{\rm 5},
    Yuchen Zhou \textsuperscript{\rm 4},
    Weiqiang Wang \textsuperscript{\rm 5},
    Shuai Lu \textsuperscript{\rm 1,}\thanks{Corresponding author}
}
\affiliations{
    \textsuperscript{\rm 1}Tsinghua University, 
    \textsuperscript{\rm 2}The University of Hong Kong, 
    \textsuperscript{\rm 3}Zhejiang University, \\
    \textsuperscript{\rm 4}National University of Singapore,
    \textsuperscript{\rm 5}Ant. Group
}

\usepackage{bibentry}
\begin{document}

\maketitle

\begin{abstract}
Retrieval-Augmented Generation (RAG) enhances large language models (LLMs) by integrating up-to-date external knowledge, yet real-world web environments present unique challenges. 
These limitations manifest as two key challenges: pervasive misinformation in the web environment, which introduces unreliable or misleading content that can degrade retrieval accuracy, and the underutilization of web tools, which, if effectively employed, could enhance query precision and help mitigate this noise, ultimately improving the retrieval results in RAG systems.
To address these issues, we propose WebFilter, a novel RAG framework that generates source-restricted queries and filters out unreliable content. This approach combines a retrieval filtering mechanism with a behavior- and outcome-driven reward strategy, optimizing both query formulation and retrieval outcomes. Extensive experiments demonstrate that WebFilter improves answer quality and retrieval precision, outperforming existing RAG methods on both in-domain and out-of-domain benchmarks.
Code is available at \textcolor{blue}{\url{https://github.com/GuoqingWang1/WebFilter}}.

\end{abstract}

\section{Introduction}

The advancement of large language models (LLMs) has driven significant progress across both industrial and academic fields~\cite{qiu2024llm, zhang2024toward, yu2025primus}. 
Despite the wide applicability of LLMs, they often struggle with knowledge-intensive queries because their knowledge can be incomplete or outdated, which leads to factual inaccuracies or hallucinations~\cite{zhang2023siren, sahoo2024large, ji2023survey}.
To address these challenges, Retrieval-Augmented Generation (RAG) enhances model performance by retrieving relevant external knowledge during inference. This approach enables the model to access up-to-date information and fill in knowledge gaps.

RAG is meant to help models access up-to-date information, but the high cost and delay of online retrieval have caused early research~\cite{chen2025research, jin2025search, song2025r1} to focus mainly on using locally stored knowledge. While these local sources are efficient, they are often outdated or incomplete, which affects model performance in real-world situations. More recent research~\cite{li2025webthinker, deepresearcher, wei2025webagentr1trainingwebagents} has started exploring RAG in web-based environments, showing the benefits of using web search during model training. However, unlike local retrieval, which relies on trusted, static data, web-based retrieval comes with its own challenges:
1) \textbf{Web environment pervasive misinformation}: the open web is saturated with misinformation, low-quality content noise~\cite{yang2025realfactbench,yang2025rama}. This significantly increases the difficulty of identifying credible sources and introduces risks of model hallucination or factual inconsistency during answer generation.
2) \textbf{Web tools underutilization}: local tools and web-based search engines differ significantly in utilization. While web tools provide \textit{advanced search operators} that help avoid outdated information and enable retrieval from trusted sources, such capabilities are unavailable in offline settings. As a result, locally trained models struggle to learn and use advanced tools, limiting their ability to filter noise and focus on reliable sources.

\begin{figure}
    \centering
    \includegraphics[width=1\linewidth]{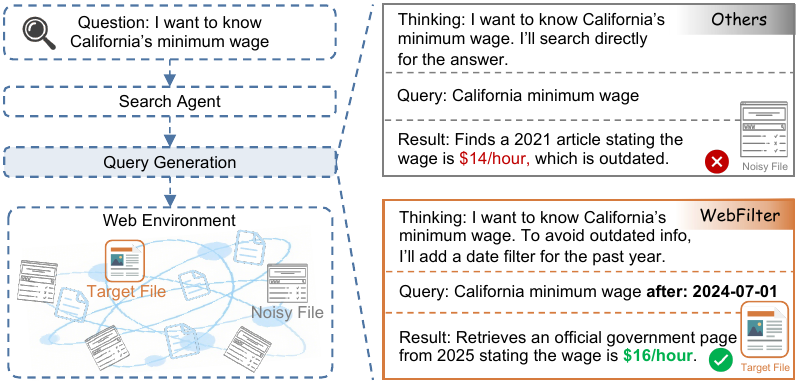}
    \caption{
    Comparison of WebFilter with Existing Methods:
    Existing methods~\cite{deepresearcher, song2025r1} often yield unreliable results in misinformation-rich web environments. WebFilter enhances accuracy by using advanced search operators to filter noise and retrieve target files.
     }
    \label{fig:main_fig}
\end{figure}

To address these challenges, we present WebFilter, a framework that improves answer accuracy by filtering noise using advanced search operators. Given the imperfections of search engines and their sensitivity to query quality (see Fig.~\ref{fig:main_fig}), our framework further enhances retrieval accuracy by formulating more effective queries and applying advanced operators. These operators, such as source selection and time filtering, enable precise retrieval by filtering out noisy sources and enhancing credibility.
However, guiding models to correctly use these operators is challenging. Our experiments show that, even with instructions, models rarely proactively use advanced operators. This is because, while existing models~\cite{li2025webthinker, deepresearcher, song2025r1} have achieved state-of-the-art results using Reinforcement Learning (RL)~\cite{kaelbling1996reinforcement}, they focus primarily on outcomes, rather than guiding behavior in web-based environments. As a result, they often rely on unreliable shortcuts. Thus, methods that effectively guide models to utilize advanced operators for noise filtering are crucial.

Therefore, to systematically integrate web search into model reasoning, we formulate retrieval as a Markov Decision Process (MDP) and guide the model to operate as an information retrieval agent capable of using advanced operators. 
To more effectively guide model behavior, we introduce an Information-Filtering Reward strategy, which combines two complementary rewards driven by both behavioral and outcome considerations. Specifically: 
(1) \textbf{Source-restricting Reward (SR)} encourages the model to proactively use advanced search operators (e.g., domain filters, date ranges), shaping query formulation strategies. In the early stages of training, SR promotes exploration even when the model's performance is suboptimal. As training progresses, SR gradually shifts focus towards accuracy, refining the model’s query formulation to prioritize precision. This transition helps balance exploration and exploitation, ensuring effective learning and improved retrieval performance.
(2) \textbf{Retrieval-precision Reward (RR)} reinforces outcomes by having a more capable, large-scale LLM evaluate the quality of retrieved content and provide feedback, enabling the model to refine its queries and improve source selection based on retrieval results.
Through the combination of structured modeling, instruction design, and reward learning, WebFilter overcomes pervasive web misinformation and fully leverages advanced web search tools for precise and reliable retrieval.
In summary, our contributions are as follows:
\begin{itemize}
    \item We propose WebFilter, a novel RAG framework explicitly designed for real-world web environments. It formulates retrieval as an MDP and trains LLMs as information retrieval agents, enabling effective mitigation of pervasive misinformation and better utilization of advanced web search tools.

    \item We introduce an information-filtering reward strategy that guides precise, source-restricted retrieval and enables robust misinformation filtering, addressing both pervasive web noise and tool underutilization.

    \item Experiments show that WebFilter achieves state-of-the-art QA performance, with advanced search operator usage rising from 10\% to 75\%, and significant gains across in-domain and out-of-domain benchmarks.

\end{itemize}

\section{Related Work}
\subsection{Agentic Retrieval Augmented Generation}
Recent work has explored agentic RAG \cite{chen2025research,jin2025search,song2025r1} to integrate retrieval into the reasoning process of LLMs. For example, methods such as ReSearch \cite{chen2025research}, Search-R1 \cite{jin2025search}, and R1-Searcher \cite{song2025r1} train LLMs to autonomously generate search queries while reasoning with a local search engine. However, LLMs trained in such local settings often struggle to generalize to real-world web environments \cite{deepresearcher}. To overcome this, methods such as WebRL \cite{qi2024webrl}, WebThinker \cite{li2025webthinker}, R1-Searcher \cite{song2025r1}, DeepResearcher \cite{deepresearcher}, WebAgent-R1 \cite{wei2025webagentr1trainingwebagents} leverage online search engines for training. Yet, compared to local settings, online environments pose greater challenges, including high API costs, network latency, and the abundance of false or redundant information, all of which hinder efficient training and retrieval \cite{deepresearcher}. 
However, due to their reliance on local web environments and the lack of source-specified retrieval data, existing reward schemes fall short in tackling real-world web challenges such as pervasive misinformation and poor use of advanced search operators.
To address this issue, we formulate retrieval as a Markov Decision Process, guided by explicit instruction on tool usage and an Information-Filtering Reward strategy, jointly enabling more structured, source-restricted querying and robust information filtering.

\subsection{Reinforcement Learning for LLMs}
Reinforcement learning (RL) has become increasingly prominent in training LLMs, supporting applications that span from preference alignment \cite{ouyang2022training, casper2023open, kaufmann2023survey} to complex tasks~\cite{hao2023reasoning,pang2024iterative, tang2025gui, xie2025dualagentadversarialframeworkrobust}.
A growing area of interest is the application of RL to tool-integrated tasks \cite{li2025torl}, which involve multi-step interactions and dynamic tool states. The high interactivity with the environment makes them a natural fit for RL. Existing research has explored RL-trained LLM agents for tool-integrated reasoning. For example, ToRL \cite{li2025torl} and Tool-N1 \cite{zhang2025nemotron} employ rule-based outcome rewards that account for both accuracy and format to guide RL, while other methods \cite{wang2025otc,sha2025sem,singh2025agentic} extend this by incorporating tool usage-based reward. 
However, most RL methods are built on local corpora without source-restricted supervision, limiting their generalization to real-world web environments with noisy information and underused search tools. We address this by formulating retrieval as an MDP and combining tool-use instruction with an Information-Filtering Reward strategy.

\section{Methodology}
\begin{figure*}[!t]
\begin{center}
\centerline{\includegraphics[width=0.97\textwidth]{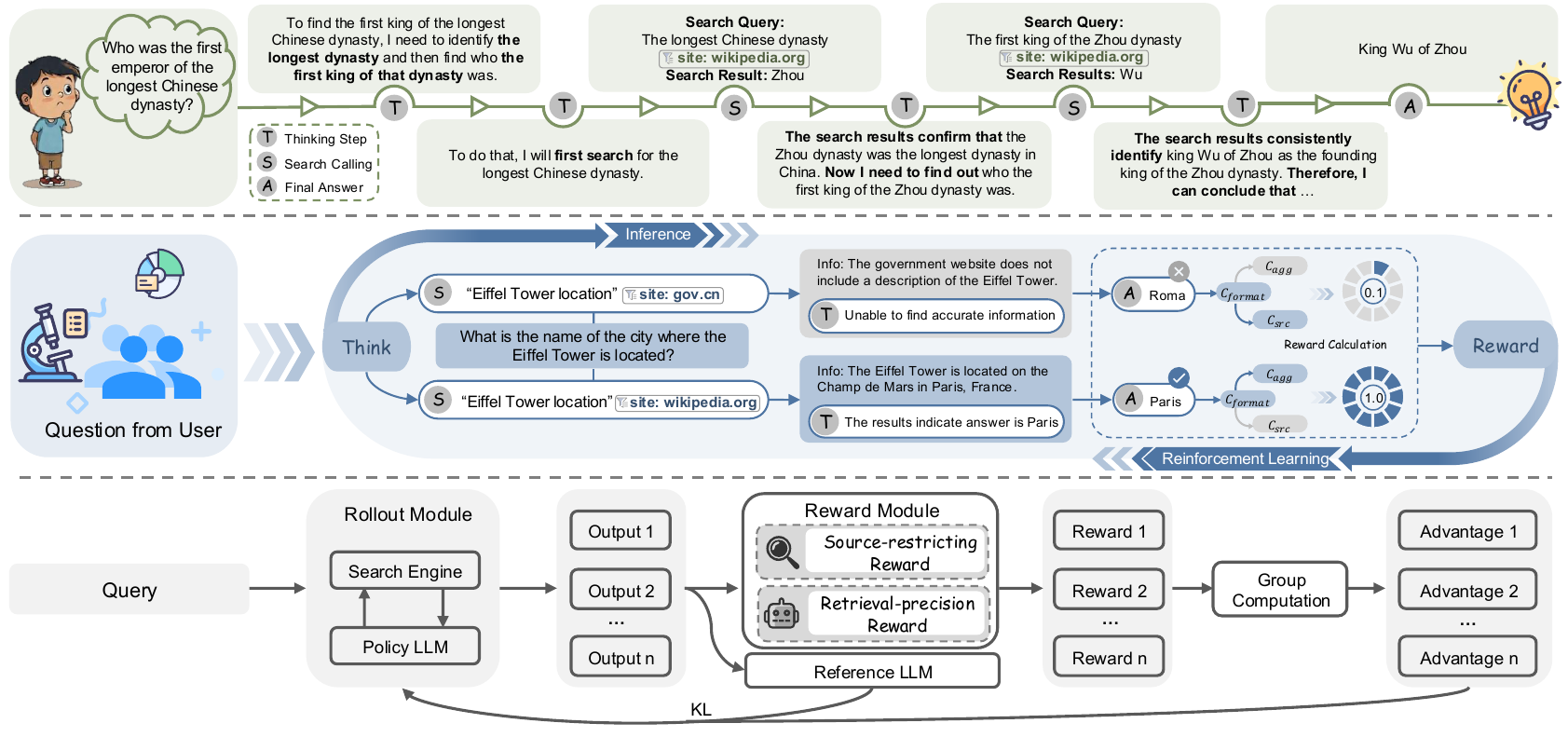}}
\caption{
Overview of the WebFilter training framework.
\textbf{Upper:} We formulate retrieval as a Markov Decision Process, where the model interacts with web search tools through step-by-step actions, including query generation and evidence selection.
\textbf{Middle:} To improve tool usage, we provide explicit instructions and demonstrations on how to issue effective, source-aware queries.
\textbf{Lower:} The policy is optimized using a behavior- and outcome-driven Information-Filtering Reward strategy, which encourages both proper tool invocation and high-quality information retrieval.
}
\label{fig:framework}
\end{center}
\end{figure*}

In this section, we introduce the WebFilter training framework, designed to enhance Retrieval-Augmented Generation (RAG) by improving query formulation and filtering unreliable web content. As shown in Fig.~\ref{fig:framework}, we model the retrieval process as a Markov Decision Process, enabling the model to decide when and how to issue search queries and integrate the retrieved information. To guide this process, we implement an Information-Filtering reward strategy, which evaluates retrieval outcomes and refines query strategies based on feedback from a stronger LLM. The following sections detail the framework and problem formulation.

\subsection{Problem Formulation}
We model the task completion as a Markov Decision Process (MDP), denoted by  \((S, A, R, \mathcal{T})\), where the state \(s_t \in S\) represents the history of previous actions at time step \(t\). At each \(t\) step, the agent selects an action \(a_t \in A\) based on the current state \(s_t\), following the policy \(\pi_\theta\). When the agent selects the "search" action (\(a_t = \text{search}\)), it updates the state by incorporating the retrieved results. Specifically, \(d_t\) refers to the content retrieved based on the search query at time step \(t\). The state transition is defined as:
\begin{equation}
s_{t+1} = T(s_t, a_t) =
\begin{cases}
[ s_t ; a_t, d_t ] & \text{if } a_t =  \text{search}, \\
[ s_t ; a_t ] & \text{otherwise}.
\end{cases}
\end{equation}
where \(T\) represents the deterministic state transition function, and the agent receives a reward \(r_t = R(s_t, a_t)\), as determined by the environment. The process terminates when the task is completed or the maximum allowed interactions are reached.

\subsection{Learning to Use Advanced Search Tools}
WebFilter is implemented as a prompt-based Information Retrieval Agent that proactively conducts web searches and reasons over retrieved results before issuing a final answer. It operates under a cost-sensitive policy, minimizing excessive queries and avoiding uncertain responses when evidence is lacking. The agent is explicitly instructed to integrate advanced search operators such as \texttt{OR}, \texttt{AND}, \texttt{NOT}, quotation marks for exact phrases, domain restrictions via \texttt{site:}, and date filters like \texttt{after:}, enabling precise, source-restricted retrieval. It also prioritizes trusted domains (e.g., \texttt{wikipedia.org}, \texttt{gov.cn}), guiding the model to generate focused, reliable, and efficient queries. 
For implementation details, please refer to our GitHub repository.

\subsection{Information-Filtering Reward Strategy}
Although the MDP formulation structures retrieval interactions, it alone cannot ensure precise query formulation or reliable filtering of web content. 
To address this, we propose an Information-Filtering Reward strategy that integrates both behavior-based and outcome-based incentives. The Source-restricting Reward (SR) acts as a behavior-based restriction, promoting the use of advanced search operators for precise, source-restricted queries. In contrast, the Retrieval-precision Reward (RR) serves as an outcome-based signal, leveraging external critique to assess and refine retrieval quality. Together, these rewards guide the model toward more effective and trustworthy web search. 
Next, we will describe its design in detail.
\paragraph{Source-restricting Reward (SR)} 
To encourage precise and source-restricted queries, we design a rule-based Source-restricting Reward that promotes the use of advanced search operators.
Specifically, the Source-restricting Reward is defined by the following formula:
\begin{equation}
\mathcal{K}=\{k_1,k_2,...,k_m\}\subseteq \Sigma^*,
\end{equation}
\begin{equation}
\mathcal{Q}=\{q_1,q_2,...,q_n\}\subseteq y,
\end{equation}
\begin{equation}
R_{src} = \mathbb{I} \left[ \exists q \in \mathcal{Q},\ \exists k \in \mathcal{K},\ ReMatch(k, q) = 1 \right],
\end{equation}
where \(\Sigma^*\) denotes the set of all sequences over the vocabulary of the policy model; \(\mathcal{K}\subseteq \Sigma^*\) is a predefined set of advanced search keyword patterns, such as ``site:", ``-", or ``AND"; \(y\subseteq \Sigma^*\) is a response generated by the policy model; and \(\mathcal{Q}\subseteq y\) is the set of search queries extracted from \(y\). The binary function \(ReMatch(k, q)\) returns 1 if query \(q\) matches the regular expression associated with pattern \(k\), and 0 otherwise. The indicator function \(\mathbb{I}[\cdot]\) equals 1 if the predicate holds and 0 otherwise. We define the Source-restricting Reward \(R_{src}\in\{0,1\}\)  to be 1 if any query in \(\mathcal{Q}\) contains an advanced search pattern from \(\mathcal{K}\), and 0 otherwise. This design explicitly promotes the use of advanced search operators in the reinforcement learning process.
\paragraph{Retrieval-precision Reward (RR)}
Beyond encouraging the use of advanced search operators, it is equally important to ensure they are applied correctly and effectively. We thus propose an LLM-based, outcome-oriented Retrieval-precision Reward (RR) that evaluates responses and provides feedback on operator correctness and quality. 
Specifically, the Retrieval-precision Reward is defined as:
\begin{equation}
    z,\, c = LLM_{Judge}(g, y,I_{Judge}),
\end{equation}
where $LLM_{Judge}$ denotes a powerful LLM used to evaluate the model's predictions, $g$ is the ground-truth answer, $y$ is the response generated by the policy model, and $I_{Judge}$ denotes the prompt template provided to $LLM_{Judge}$, as shown in \ref{Metrics}. The scalar $z \in \{0,1\}$ measures the correctness of the predicted answer, while $c \in \{0,1\}$ evaluates whether the use of advanced search syntax contributed to the retrieval quality. 

\begin{table*}[h]
\centering
\caption{
Performance of different methods on in-domain datasets, evaluated with rule-based ($ACC_R$) and LLM-based ($ACC_L$) metrics. Best results are highlighted in bold.
}
\label{tab:in_domain_performance}

{\small  
\setlength{\tabcolsep}{6pt}  
\begin{tabular}{
    ll
    S[table-format=2.1] S[table-format=2.1]
    S[table-format=2.1] S[table-format=2.1]
    S[table-format=2.1] S[table-format=2.1]
    S[table-format=2.1] S[table-format=2.1]
}
\toprule
\textbf{Environment} & \textbf{Method} 
& \multicolumn{2}{c}{\textbf{NQ}} 
& \multicolumn{2}{c}{\textbf{TQ}} 
& \multicolumn{2}{c}{\textbf{HotpotQA}} 
& \multicolumn{2}{c}{\textbf{2Wiki}} \\ 
\cmidrule(lr){3-4}
\cmidrule(lr){5-6}
\cmidrule(lr){7-8}
\cmidrule(lr){9-10}
& & {\small $ACC_R$} & {\small $ACC_L$} 
  & {\small $ACC_R$} & {\small $ACC_L$} 
  & {\small $ACC_R$} & {\small $ACC_L$} 
  & {\small $ACC_R$} & {\small $ACC_L$} \\ 
\midrule
\multirow{3}{*}{Direct Reasoning} 
& Qwen3-32B          & 16.5 & 36.3 & 11.6 & 52.9 & 29.9 & 19.1 & 26.5 & 19.9 \\
& Gemini-2.0-Flash   & 20.7 & 47.1 & 16.2 & 68.9 & 42.4 & 29.9 & 32.5 & 23.8 \\
& GPT-4o              & 23.1 & 53.9 & 17.6 & 73.2 & 50.2 & 37.1 & 38.6 & 30.5 \\
\midrule
\multirow{3}{*}{Local RAG} 
& Search-o1          & 34.5 & 57.4 & 52.6 & 61.1 & 31.6 & 40.8 & 28.6 & 32.8 \\
& Search-r1-base     & 45.4 & 60.0 & 71.9 & 76.2 & 55.9 & 63.0 & 44.6 & 47.9 \\
& Search-r1-instruct & 33.1 & 49.6 & 44.7 & 49.2 & 45.7 & 52.5 & 43.4 & 48.8 \\
\midrule
\multirow{3}{*}{Web Search} 
& R1-Searcher        & 35.4 & 52.3 & 73.1 & 79.1 & 44.8 & 53.1 & 59.4 & 65.8 \\
& DeepResearcher     & 39.6 & 61.9 & \textbf{78.4} & 85.0 & 52.8 & 64.3 & 59.7 & 66.6 \\
& \textbf{WebFilter (Ours)} 
                     & \textbf{40.1} & \textbf{63.1} 
                     & 77.5 & \textbf{85.4} 
                     & \textbf{55.1} & \textbf{65.2} 
                     & \textbf{60.1} & \textbf{67.5} \\
\bottomrule
\end{tabular}
}
\end{table*}

\paragraph{Reward Aggregation} 
The final reward $R$ is computed as:
\begin{equation}
    R = 
    \begin{cases}
        -1 & \text{if } \neg C_{\text{format}}, \\[6pt]
        f(z, c, R_{f1}) & \text{if } C_{\text{format}} \wedge C_{\text{agg}}, \\[6pt]
        0.1 & \text{if } C_{\text{format}} \wedge \neg C_{\text{agg}} \wedge C_{\text{src}}, \\[6pt]
        0 & \text{otherwise}.
    \end{cases}
\end{equation}
Here, $C_{\text{format}}$ denotes correct output format; 
$C_{\text{agg}}$ indicates $f(R_{llm}, R_{f1}) \neq 0$; 
$C_{\text{src}}$ means $R_{src} = 1$.
The F1 Reward $R_{f1}$ is computed as:
\begin{equation}
    R_{f1} = \frac{2 \times IN}{PN + RN},
\end{equation}
where $PN$ is the word count of the predicted answer, $RN$ is the word count of the reference answer, and $IN$ is the word count of overlapping words between them.
The aggregation function $f(\cdot)$ is defined as:
\begin{equation}
    f(u,v,w) = \alpha u + \beta v + (1-\alpha-\beta) w,
\end{equation}
where $\alpha$ and $\beta$ are hyperparameters used to balance the influence of each reward on the policy optimization process.

\subsection{RL Training Framework}
\paragraph{Policy Optimization} 
In this work, we adopt the Group Relative Policy Optimization (GRPO) \cite{shao2024deepseekmath}, which improves the current policy $\pi_{\theta}$ by leveraging a reference policy $\pi_{\theta_{\text{ref}}}$ and a set of rollouts generated by a old policy $\pi_{\theta_{\text{old}}}$. To support search engine calls, the GRPO objective is extended as follows:

{\footnotesize
\begin{equation}
\begin{aligned}
\mathcal{J}(\theta) =\ & \mathbb{E}_{x \sim \mathcal{D},\, \{y_i\}_{i=1}^{G} \sim \pi_{\theta_{\text{old}}}(\cdot|x)} \Bigg[ \frac{1}{G} \sum_{i=1}^{G} \min \Bigg( \frac{\pi_{\theta}(y_i|x)}{\pi_{\theta_{\text{old}}}(y_i|x)} A_i, \\[4pt]
& \quad\text{clip} \left( \frac{\pi_{\theta}(y_i|x)}{\pi_{\theta_{\text{old}}}(y_i|x)},\ 1 - \epsilon,\ 1 + \epsilon \right) A_i \Bigg) \\
& \quad- \beta\, \mathbb{D}_{\mathrm{KL}}\left(\pi_{\theta} \,\|\, \pi_{\theta_{\mathrm{ref}}}\right) \Bigg].
\end{aligned}
\end{equation}
}

Here, $x$ denotes an input sampled from the data distribution $\mathcal{D}$, and $y_i$ represents a trajectory generated by $\pi_{\theta_{\text{old}}}$. $\mathbb{D}_{\mathrm{KL}}$ is the estimated KL divergence \cite{shao2024deepseekmath}, and $\epsilon$, $\beta$ are hyperparameters that control the trust region and regularization strength, respectively.
The reward $r_i$ for each $y_i$ is computed jointly across trajectories:
\begin{equation}
    r_1, r_2, \ldots, r_G = R(y_1, y_2, \ldots, y_G),
\end{equation}
and the advantage $A_i$ is normalized within the batch as:
\begin{equation}
    A_i = \frac{r_i - \text{mean}(r_1, r_2, \ldots, r_G)}{\text{std}(r_1, r_2, \ldots, r_G)}.
\end{equation}

This objective encourages stable policy improvement, enabling effective integration of retrieval-based reasoning into the learning process.

\section{Experiments}
\subsection{Benchmarks}
Our experimental setting is built on question answering datasets that assess reasoning and retrieval capabilities in diverse scenarios. 
For in-domain evaluation, we use the development sets of Natural Questions~(NQ)~\cite{kwiatkowski2019natural}, TriviaQA~(TQ)~\cite{joshi2017triviaqa}, HotpotQA \cite{yang2018hotpotqa}, and 2Wiki \cite{ho2020constructing}. 
For out-of-domain evaluation, we include the complex open-domain reasoning dataset MuSiQue \cite{trivedi2022musique} and the web-search-focused benchmark Bamboogle \cite{press2022measuring}, which differ in question style and information distribution.
To ensure a balanced and consistent evaluation across datasets, we select a fixed number of examples from each. Specifically, 512 examples are chosen from the development sets of NQ, TQ, HotpotQA, 2Wiki, and MuSiQue, and all 125 examples are selected from the development set of Bamboogle.

\subsection{{Baselines}}
To evaluate WebFilter's effectiveness, we compare it against several baselines representing different methodologies:

\noindent \textbf{Direct Reasoning}: Models relying solely on internal knowledge, such as Qwen3-32B~\cite{yang2025qwen3}, Gemini-2.0-Flash~\cite{team2023gemini}, and GPT-4o~\cite{hurst2024gpt}.

\noindent \textbf{Local RAG}: Methods retrieving knowledge from offline documents. For example, Search-o1~\cite{li2025search} performs multi-step reasoning by generating search queries and using the retrieved snippets as context, while Search-r1-base~\cite{jin2025search} retrieves evidence from Wikipedia during both training and inference. Search-r1-instruct~\cite{jin2025search} differs by initializing the actor with an instruct-tuned language model to guide the retrieval process.

\noindent \textbf{Web Search}: Methods utilizing online tools. Both our approach and R1-Searcher~\cite{song2025r1} rely on the Google API for web search. In addition to Google search, DeepResearcher~\cite{deepresearcher} integrates a WebBrowser tool for web navigation, which leads to increased time spent browsing and accessing websites, thereby slowing down the overall training speed. All methods, including ours, employ Qwen-2.5-7B-instruct~\cite{yang2025qwen25} for model inference.

\subsection{Metrics}
\label{Metrics}
We evaluate model performance using both rule-based ($ACC_R$) and LLM-based ($ACC_L$) metrics. The rule-based metric uses an F1 score to measure overlap between predictions and reference answers, reflecting factual precision. For $ACC_L$, we adopt the LLM-as-Judge framework \cite{zheng2023judging}, where GPT-4o-mini \cite{hurst2024gpt} assesses whether model answers align semantically with the references, thus capturing nuances beyond exact matching.

\subsection{Implementation Details}
We implement our model using the VeRL framework~\cite{verl} and adopt Qwen2.5-7B-Instruct~\cite{yang2025qwen25} as the backbone. The hyperparameters for the aggregation function are set as $\alpha = 0.4$ and $\beta = 0.2$. The learning rate is set to 1e-5, and training proceeds with a mini-batch size of 4,096. Each iteration processes 256 samples, generating 16 rollouts per sample. Additionally, we apply a sampling temperature of 1.0 and limit the maximum retrieval count to 10. We apply loss masking to update only model-generated tokens. 
Our Retrieval-precision Reward uses Qwen3-30B-A3B~\cite{yang2025qwen3} as the judge model, which is free and can be deployed locally.

\begin{table}[!t]
\centering
\footnotesize
\caption{
Performance of methods on out-of-domain datasets.
}
\label{tab:ood_performance}
\begin{tabular}{
    l
    S[table-format=2.1] S[table-format=2.1]
    S[table-format=2.1] S[table-format=2.1]
}
\toprule
\textbf{Method} 
& \multicolumn{2}{c}{\textbf{Musique}} 
& \multicolumn{2}{c}{\textbf{Bamboogle}} \\ 
\cmidrule(lr){2-3}
\cmidrule(lr){4-5}
& {\scriptsize $ACC_R$} & {\scriptsize $ACC_L$} 
  & {\scriptsize $ACC_R$} & {\scriptsize $ACC_L$} \\
\midrule
Qwen3-32B         & 10.7 & 4.9  & 24.7 & 18.4 \\
Gemini-2.0-Flash  & 11.4 & 6.1  & 36.5 & 28.0 \\
GPT-4o             & 22.5 & 15.0 & 52.6 & 43.2 \\
\midrule
Search-o1          & 16.8 & 21.3 & 46.6 & 53.6 \\
Search-r1-base     & 26.7 & 27.5 & 56.6 & 57.6 \\
Search-r1-instruct & 26.5 & 28.3 & 45.0 & 47.2 \\
\midrule
R1-Searcher          & 22.8 & 25.6 & 64.8 & 65.6 \\
DeepResearcher       & \textbf{27.1} & 29.3 & 71.0 & 72.8 \\
\textbf{WebFilter (Ours)} & 24.5 & \textbf{30.0} & \textbf{73.1} & \textbf{74.3} \\
\bottomrule
\end{tabular}
\end{table}

\begin{table*}[!t]
\centering
\caption{
Performance of different WebFilter variants across in-domain datasets (NQ, TQ, HotpotQA, 2Wiki) and out-of-domain datasets (Musique, Bamboogle). 
“SR” denotes the Source-restricting Reward, and “RR” denotes the Retrieval-precision Reward.
}
\label{tab:webfilter_variants}

{\footnotesize  % 使用更小的字体
\setlength{\tabcolsep}{3pt}  % 减小列间距
\begin{tabular}{l
                r r
                r r
                r r
                r r
                r r
                r r}
\toprule
\textbf{Methods} 
& \multicolumn{2}{c}{\textbf{NQ}} 
& \multicolumn{2}{c}{\textbf{TQ}} 
& \multicolumn{2}{c}{\textbf{HotpotQA}} 
& \multicolumn{2}{c}{\textbf{2Wiki}} 
& \multicolumn{2}{c}{\textbf{Musique}} 
& \multicolumn{2}{c}{\textbf{Bamboogle}} \\ 
\cmidrule(lr){2-3}
\cmidrule(lr){4-5}
\cmidrule(lr){6-7}
\cmidrule(lr){8-9}
\cmidrule(lr){10-11}
\cmidrule(lr){12-13}
& {\footnotesize $ACC_R$} & {\footnotesize $ACC_L$} 
& {\footnotesize $ACC_R$} & {\footnotesize $ACC_L$} 
& {\footnotesize $ACC_R$} & {\footnotesize $ACC_L$} 
& {\footnotesize $ACC_R$} & {\footnotesize $ACC_L$} 
& {\footnotesize $ACC_R$} & {\footnotesize $ACC_L$} 
& {\footnotesize $ACC_R$} & {\footnotesize $ACC_L$} \\
\midrule
Base 
& 41.2 & 63.6 
& 78.5  & 82.6 
& 49.4  & 59.9 
& 55.2  & 58.2 
& 22.7  & 26.2 
& 64.9  & 65.8 \\
Base+SR 
& \textbf{41.4} & \textbf{64.3} 
& \textbf{79.0} & \textbf{86.0} 
& 50.6 & 60.1 
& 59.7 & 65.3 
& 24.5 & 27.6 
& 64.6 & 65.2 \\
\textbf{Base+SR+RR(Ours)}
& 40.1 & 63.1 
& 77.5 & 85.4 
& \textbf{55.1} & \textbf{65.2} 
& \textbf{60.1} & \textbf{67.5} 
& \textbf{24.5} & \textbf{29.0} 
& \textbf{73.1} & \textbf{74.3} \\
\bottomrule
\end{tabular}
}
\end{table*}

\subsection{Results on In-Domain Settings}
WebFilter achieves state-of-the-art performance across all four in-domain datasets, demonstrating notable strengths in multi-hop reasoning tasks, as shown in Tab.~\ref{tab:in_domain_performance}. 
On HotpotQA, it outperforms DeepResearcher by 2.3\% in $ACC_R$, despite DeepResearcher relying on a browser tool with higher latency for broader web exploration.
This advantage arises from WebFilter’s ability to formulate precise, source-restricted queries using advanced search operators, effectively reducing noise in retrieved documents. Compared to local RAGs, the performance gap on 2Wiki becomes more pronounced, with WebFilter achieving around a 17\% higher $ACC_R$ through selective access to trusted external sources.
These improvements reflect deliberate design choices. 
Unlike methods limited to fixed domains, such as R1-Searcher, which restricts access to Wikipedia, or those reliant on extensive web browsing, which introduces higher latency (e.g., DeepResearcher), WebFilter focuses on generating precise, source-restricted queries for unrestricted Google API searches.
This strategy strikes a balance between retrieval flexibility and high precision, minimizing irrelevant content and enhancing the quality of retrieved contexts. As indicated by gains in $ACC_L$, WebFilter retrieves evidence that aligns more closely with ground-truth answers, thereby supporting stronger reasoning.

\subsection{Results on Out-of-Domain Settings}
WebFilter consistently shows strong generalization on out-of-domain datasets. As shown in Tab.~\ref{tab:ood_performance}, WebFilter achieves the highest $ACC_L$ on the challenging open-domain Musique dataset (30.0\%) and the web-search-heavy Bamboogle dataset (74.3\%). These results suggest that WebFilter retrieves evidence more semantically aligned with ground-truth answers, which is critical for reasoning beyond exact matching.
WebFilter also maintains competitive $ACC_R$ scores, particularly on Bamboogle (73.1\%), indicating its ability to preserve factual consistency in new domains. Moreover, it outperforms DeepResearcher on $ACC_L$ for Musique (30.0\% vs. 29.3\%), demonstrating its strength in handling difficult open-domain questions.
WebFilter’s ability to retrieve semantically relevant and factually consistent information across diverse topics underscores its practical value for real-world applications involving domain shifts and open-ended queries.

\begin{figure}[]
    \centering
    \includegraphics[width=0.9\linewidth]{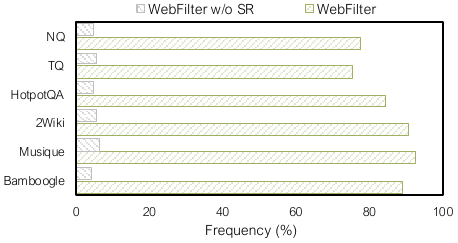}
    \vspace{-2mm}
    \caption{Frequency of advanced operators across variants.}
    \label{fig:frequency}
\end{figure}

\begin{figure}[h]
    \centering
    \subfloat[$ACC_R$ over training steps]{%
        \vspace{-3mm}
        \includegraphics[width=0.95\linewidth]{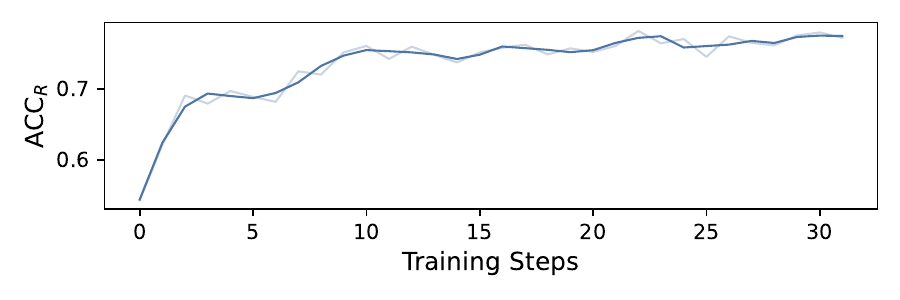}%
        \label{fig:f1}%
    }\\
    \vspace{-1mm}
    \subfloat[Tool call counts over training steps]{%
        \vspace{-3mm}
        \includegraphics[width=0.95\linewidth]{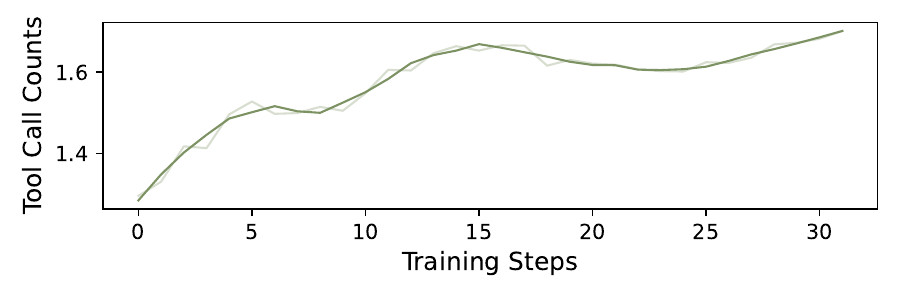}%
        \label{fig:tool}%
    }\\
    \vspace{-1mm}
    \subfloat[Response length over training steps]{%
        \vspace{-3mm}
        \includegraphics[width=0.95\linewidth]{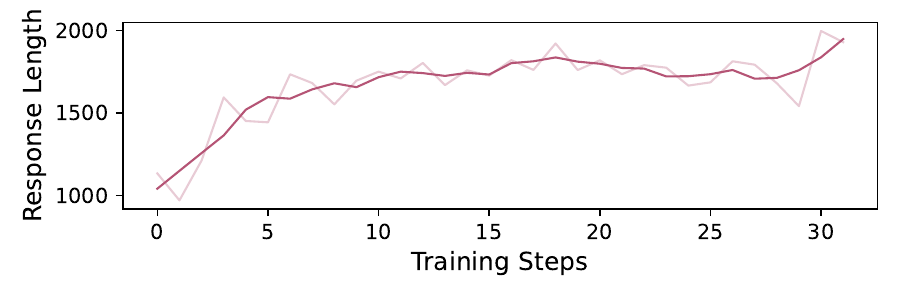}%
        \label{fig:resp}%
    }
    \vspace{-1mm}
    \caption{Training dynamics showing (a) QA accuracy ($ACC_R$), (b) tool call behavior, and (c) response length evolution across training steps.}
    \label{fig:training}
\end{figure}

\begin{figure*}
    \centering
    \includegraphics[width=1\linewidth]{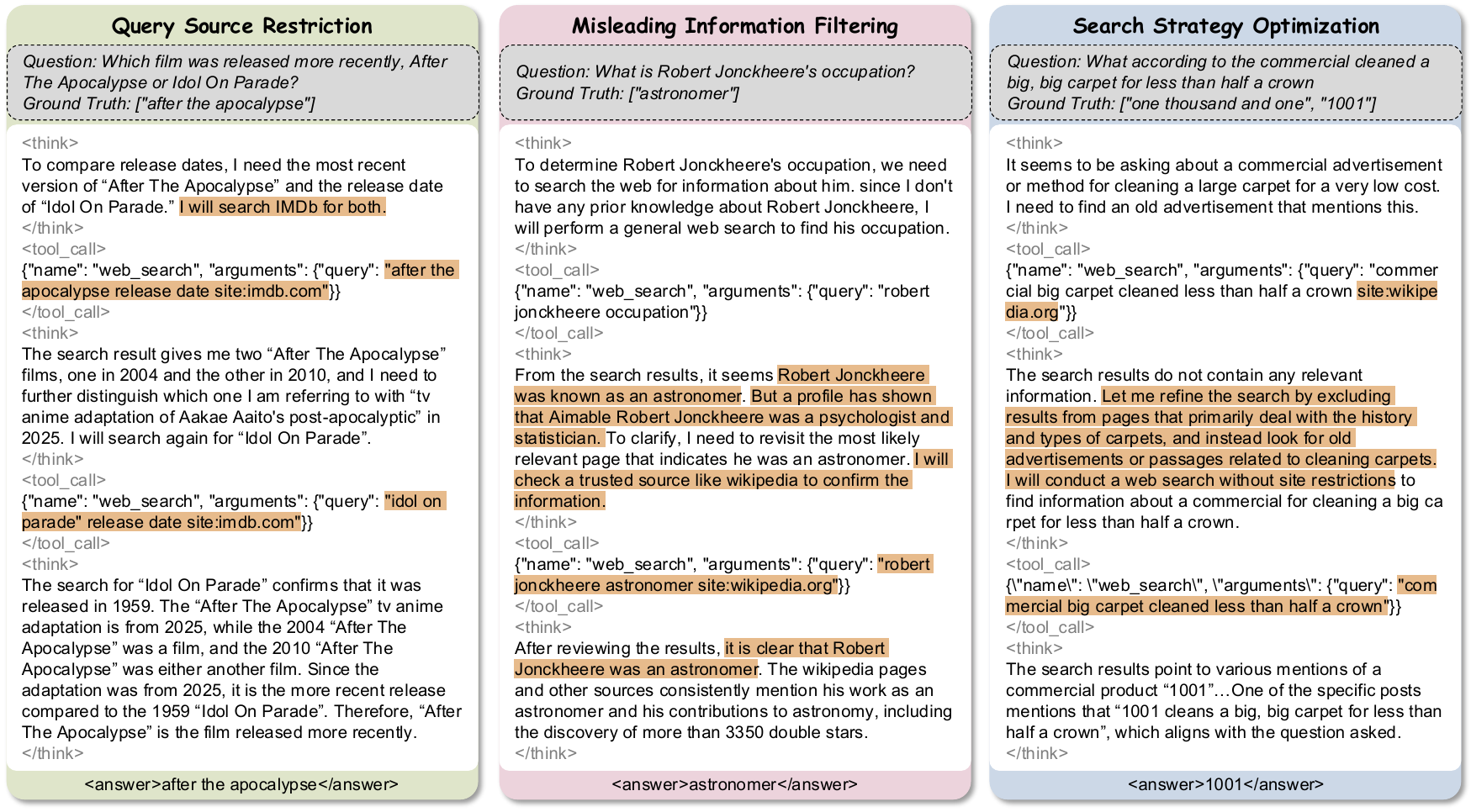}
    \caption{Case studies showing how WebFilter improves QA by (a) narrowing searches to authoritative sources for precise results, (b) verifying ambiguous or conflicting information via trusted sites, and (c) adaptively refining search queries when initial attempts are insufficient.
}
    \label{fig:case_study}
\end{figure*}

\subsection{Ablation Study}
Tab.~\ref{tab:webfilter_variants} shows the performance of WebFilter variants on both in-domain datasets (NQ, TQ, HotpotQA, 2Wiki) and out-of-domain datasets (Musique, Bamboogle). 
From the results, we observe that incorporating the Source-restricting Reward (SR) leads to noticeable gains on in-domain datasets, improving the model's ability to retrieve information from reliable sources. SR also encourages the use of advanced operators. As shown in Fig.~\ref{fig:frequency}, SR-guided queries include advanced operators more than 75\% of the time, compared to less than 10\% without SR.
Adding the Retrieval-precision Reward (RR) further boosts performance, especially on out-of-domain datasets. RR refines the retrieval process, aligning the generated evidence more closely with ground-truth answers. For instance, on Bamboogle, $ACC_R$ improves by 8.5\%, from 64.6\% to 73.1\%, and $ACC_L$ rises from 65.2\% to 74.3\%. This improvement is mainly due to Bamboogle's focus on web-search-based reasoning, where WebFilter’s ability to leverage advanced search operators and filter irrelevant content enhances retrieval of high-quality, relevant evidence.
The combination of SR and RR yields the best results, with SR boosting retrieval precision and RR enhancing generalization. These findings demonstrate the effectiveness of our reward framework in improving both retrieval quality and reasoning performance.

\subsection{Training Dynamics Analysis}
Fig.~\ref{fig:training} illustrates the training dynamics of our model, focusing on both QA performance and behavioral changes. As shown in Fig.~\ref{fig:f1}, $ACC_R$ steadily increases throughout training, rising from approximately 0.52 to around 0.77 on the TQ dataset. This indicates continuous improvements in retrieval accuracy as the model progresses.

To capture broader trends, Fig.~\ref{fig:tool} and Fig.~\ref{fig:resp} show the average tool call counts and response lengths across all evaluated datasets. In Fig.~\ref{fig:tool}, tool call counts increase during early training, plateau around step 20, and then rise again. This pattern suggests that the model gradually incorporates more frequent retrieval as training progresses. Meanwhile, Fig.~\ref{fig:resp} reveals that the average response length grows from about 1,000 tokens to nearly 2,000 tokens, indicating that the model generates more detailed and comprehensive responses over time.
Overall, these results demonstrate that our model not only improves retrieval accuracy but also adapts its tool usage and response strategies, leading to more effective and informative outputs.

\subsection{Case Study}
We present three representative cases in Fig.~\ref{fig:case_study} to illustrate WebFilter’s intelligent retrieval behaviors. Through systematic analysis, we identify three key behavioral patterns that define the model’s advanced search capabilities:

\noindent \textbf{Query Source Restriction.} For domain-specific queries, WebFilter proactively limits search scopes to authoritative sources. In Case 1, when asked about film release dates, it automatically appends “site:imdb.com” to queries. This targeted approach ensures precise, trustworthy results while reducing latency by filtering out irrelevant information.

\noindent \textbf{Misleading Information Filtering.} WebFilter demonstrates strong disambiguation skills for resolving conflicting information. In Case 2, when the initial search reveals both an astronomer and a similarly named psychologist, the model detects the ambiguity and performs a refined follow-up search limited to Wikipedia, successfully verifying the correct occupation.

\noindent \textbf{Search Strategy Optimization.} WebFilter dynamically adjusts its retrieval strategy when initial searches are insufficient. In Case 3, failing to find relevant information via a Wikipedia-restricted query, the model expands its search without site constraints, ultimately locating references to the product “1001,” which matches the question context.

These cases highlight WebFilter’s ability to reason about search scope, verify information across sources, and adapt strategies to improve retrieval effectiveness.

\section{Limitations}
While our approach shows progress, it has some limitations. To enhance RAG capabilities, improving search quality is crucial, but not sufficient on its own. 
In many cases, errors occur not because the model fails to retrieve relevant information, but because it struggles to correctly interpret and reason with the retrieved data.
Our model, when combined with improvements in reasoning abilities, can deliver even greater value in future work.

\section{Conclusion}
We present WebFilter, a framework that improves Retrieval-Augmented Generation~(RAG) by leveraging advanced search operators for precise, source-aware queries and misinformation filtering. By modeling retrieval as a Markov Decision Process, WebFilter learns to effectively use web search tools. Experiments demonstrate strong gains on both in-domain and out-of-domain QA. Future work will explore broader web interactions to further enhance real-world RAG performance.

\section*{Acknowledgements}
This work was supported by the Ant Group Research Intern Program.

\bibliography{aaai2026}

@misc{wei2025webagentr1trainingwebagents,
      title={WebAgent-R1: Training Web Agents via End-to-End Multi-Turn Reinforcement Learning}, 
      author={Zhepei Wei and Wenlin Yao and Yao Liu and Weizhi Zhang and Qin Lu and Liang Qiu and Changlong Yu and Puyang Xu and Chao Zhang and Bing Yin and Hyokun Yun and Lihong Li},
      year={2025},
      eprint={2505.16421},
      archivePrefix={arXiv},
      primaryClass={cs.CL},
      url={https://arxiv.org/abs/2505.16421}, 
}

@article{verl,
  title   = {HybridFlow: A Flexible and Efficient RLHF Framework},
  author  = {Sheng, Guangming and Zhang, Chi and Ye, Zilingfeng and ...},
  journal = {arXiv preprint arXiv:2409.19256},
  year    = {2024}
}

@inproceedings{ho2020constructing,
  title={Constructing A Multi-hop QA Dataset for Comprehensive Evaluation of Reasoning Steps},
  author={Ho, Xanh and Nguyen, Anh-Khoa Duong and Sugawara, Saku and Aizawa, Akiko},
  booktitle={Proceedings of the 28th International Conference on Computational Linguistics},
  pages={6609--6625},
  year={2020}
}

@article{hurst2024gpt,
  title={Gpt-4o system card},
  author={Hurst, Aaron and Lerer, Adam and Goucher, Adam P and Perelman, Adam and Ramesh, Aditya and Clark, Aidan and Ostrow, AJ and Welihinda, Akila and Hayes, Alan and Radford, Alec and others},
  journal={arXiv preprint arXiv:2410.21276},
  year={2024}
}

@article{zheng2023judging,
  title={Judging llm-as-a-judge with mt-bench and chatbot arena},
  author={Zheng, Lianmin and Chiang, Wei-Lin and Sheng, Ying and Zhuang, Siyuan and Wu, Zhanghao and Zhuang, Yonghao and Lin, Zi and Li, Zhuohan and Li, Dacheng and Xing, Eric and others},
  journal={Advances in Neural Information Processing Systems},
  volume={36},
  pages={46595--46623},
  year={2023}
}

@article{press2022measuring,
  title={Measuring and narrowing the compositionality gap in language models},
  author={Press, Ofir and Zhang, Muru and Min, Sewon and Schmidt, Ludwig and Smith, Noah A and Lewis, Mike},
  journal={arXiv preprint arXiv:2210.03350},
  year={2022}
}

@article{trivedi2022musique,
  title={MuSiQue: Multihop Questions via Single-hop Question Composition},
  author={Trivedi, Harsh and Balasubramanian, Niranjan and Khot, Tushar and Sabharwal, Ashish},
  journal={Transactions of the Association for Computational Linguistics},
  volume={10},
  pages={539--554},
  year={2022},
  publisher={MIT Press One Broadway, 12th Floor, Cambridge, Massachusetts 02142, USA~…}
}

@article{yang2018hotpotqa,
  title={HotpotQA: A dataset for diverse, explainable multi-hop question answering},
  author={Yang, Zhilin and Qi, Peng and Zhang, Saizheng and Bengio, Yoshua and Cohen, William W and Salakhutdinov, Ruslan and Manning, Christopher D},
  journal={arXiv preprint arXiv:1809.09600},
  year={2018}
}

@article{joshi2017triviaqa,
  title={Triviaqa: A large scale distantly supervised challenge dataset for reading comprehension},
  author={Joshi, Mandar and Choi, Eunsol and Weld, Daniel S and Zettlemoyer, Luke},
  journal={arXiv preprint arXiv:1705.03551},
  year={2017}
}

@article{kwiatkowski2019natural,
  title={Natural questions: a benchmark for question answering research},
  author={Kwiatkowski, Tom and Palomaki, Jennimaria and Redfield, Olivia and Collins, Michael and Parikh, Ankur and Alberti, Chris and Epstein, Danielle and Polosukhin, Illia and Devlin, Jacob and Lee, Kenton and others},
  journal={Transactions of the Association for Computational Linguistics},
  volume={7},
  pages={453--466},
  year={2019},
  publisher={MIT Press One Rogers Street, Cambridge, MA 02142-1209, USA journals-info~…}
}

@article{singh2025agentic,
  title={Agentic reasoning and tool integration for llms via reinforcement learning},
  author={Singh, Joykirat and Magazine, Raghav and Pandya, Yash and Nambi, Akshay},
  journal={arXiv preprint arXiv:2505.01441},
  year={2025}
}

@article{sha2025sem,
  title={SEM: Reinforcement Learning for Search-Efficient Large Language Models},
  author={Sha, Zeyang and Cui, Shiwen and Wang, Weiqiang},
  journal={arXiv preprint arXiv:2505.07903},
  year={2025}
}

@article{wang2025otc,
  title={Otc: Optimal tool calls via reinforcement learning},
  author={Wang, Hongru and Qian, Cheng and Zhong, Wanjun and Chen, Xiusi and Qiu, Jiahao and Huang, Shijue and Jin, Bowen and Wang, Mengdi and Wong, Kam-Fai and Ji, Heng},
  journal={arXiv e-prints},
  pages={arXiv--2504},
  year={2025}
}

@article{zhang2025nemotron,
  title={Nemotron-research-tool-n1: Tool-using language models with reinforced reasoning},
  author={Zhang, Shaokun and Dong, Yi and Zhang, Jieyu and Kautz, Jan and Catanzaro, Bryan and Tao, Andrew and Wu, Qingyun and Yu, Zhiding and Liu, Guilin},
  journal={arXiv preprint arXiv:2505.00024},
  year={2025}
}

@article{li2025torl,
  title={Torl: Scaling tool-integrated rl},
  author={Li, Xuefeng and Zou, Haoyang and Liu, Pengfei},
  journal={arXiv preprint arXiv:2503.23383},
  year={2025}
}

@article{pang2024iterative,
  title={Iterative reasoning preference optimization},
  author={Pang, Richard Yuanzhe and Yuan, Weizhe and He, He and Cho, Kyunghyun and Sukhbaatar, Sainbayar and Weston, Jason},
  journal={Advances in Neural Information Processing Systems},
  volume={37},
  pages={116617--116637},
  year={2024}
}

@article{hao2023reasoning,
  title={Reasoning with language model is planning with world model},
  author={Hao, Shibo and Gu, Yi and Ma, Haodi and Hong, Joshua Jiahua and Wang, Zhen and Wang, Daisy Zhe and Hu, Zhiting},
  journal={arXiv preprint arXiv:2305.14992},
  year={2023}
}

@article{kaufmann2023survey,
  title={A survey of reinforcement learning from human feedback},
  author={Kaufmann, Timo and Weng, Paul and Bengs, Viktor and H{\"u}llermeier, Eyke},
  journal={arXiv preprint arXiv:2312.14925},
  volume={10},
  year={2023}
}

@article{casper2023open,
  title={Open problems and fundamental limitations of reinforcement learning from human feedback},
  author={Casper, Stephen and Davies, Xander and Shi, Claudia and Gilbert, Thomas Krendl and Scheurer, J{\'e}r{\'e}my and Rando, Javier and Freedman, Rachel and Korbak, Tomasz and Lindner, David and Freire, Pedro and others},
  journal={arXiv preprint arXiv:2307.15217},
  year={2023}
}

@article{ouyang2022training,
  title={Training language models to follow instructions with human feedback},
  author={Ouyang, Long and Wu, Jeffrey and Jiang, Xu and Almeida, Diogo and Wainwright, Carroll and Mishkin, Pamela and Zhang, Chong and Agarwal, Sandhini and Slama, Katarina and Ray, Alex and others},
  journal={Advances in neural information processing systems},
  volume={35},
  pages={27730--27744},
  year={2022}
}

@article{ji2023survey,
  title={Survey of hallucination in natural language generation},
  author={Ji, Ziwei and Lee, Nayeon and Frieske, Rita and Yu, Tiezheng and Su, Dan and Xu, Yan and Ishii, Etsuko and Bang, Ye Jin and Madotto, Andrea and Fung, Pascale},
  journal={ACM computing surveys},
  volume={55},
  number={12},
  pages={1--38},
  year={2023},
  publisher={ACM New York, NY}
}

@article{sahoo2024large,
  title={Large language models for biomedicine: foundations, opportunities, challenges, and best practices},
  author={Sahoo, Satya S and Plasek, Joseph M and Xu, Hua and Uzuner, {\"O}zlem and Cohen, Trevor and Yetisgen, Meliha and Liu, Hongfang and Meystre, St{\'e}phane and Wang, Yanshan},
  journal={Journal of the American Medical Informatics Association},
  volume={31},
  number={9},
  pages={2114--2124},
  year={2024},
  publisher={Oxford University Press}
}

@article{zhang2023siren,
  title={Siren's song in the AI ocean: a survey on hallucination in large language models},
  author={Zhang, Yue and Li, Yafu and Cui, Leyang and Cai, Deng and Liu, Lemao and Fu, Tingchen and Huang, Xinting and Zhao, Enbo and Zhang, Yu and Chen, Yulong and others},
  journal={arXiv preprint arXiv:2309.01219},
  year={2023}
}

@article{yu2025primus,
  title={Primus: A pioneering collection of open-source datasets for cybersecurity LLM training},
  author={Yu, Yao-Ching and Chiang, Tsun-Han and Tsai, Cheng-Wei and Huang, Chien-Ming and Tsao, Wen-Kwang},
  journal={arXiv preprint arXiv:2502.11191},
  year={2025}
}

@inproceedings{zhang2024toward,
  title={Toward mitigating misinformation and social media manipulation in llm era},
  author={Zhang, Yizhou and Sharma, Karishma and Du, Lun and Liu, Yan},
  booktitle={Companion Proceedings of the ACM Web Conference 2024},
  pages={1302--1305},
  year={2024}
}

@article{qiu2024llm,
  title={LLM-based agentic systems in medicine and healthcare},
  author={Qiu, Jianing and Lam, Kyle and Li, Guohao and Acharya, Amish and Wong, Tien Yin and Darzi, Ara and Yuan, Wu and Topol, Eric J},
  journal={Nature Machine Intelligence},
  volume={6},
  number={12},
  pages={1418--1420},
  year={2024},
  publisher={Nature Publishing Group}
}

@article{tang2025gui,
  title={GUI-G: Gaussian Reward Modeling for GUI Grounding},
  author={Tang, Fei and Gu, Zhangxuan and Lu, Zhengxi and Liu, Xuyang and Shen, Shuheng and Meng, Changhua and Wang, Wen and Zhang, Wenqi and Shen, Yongliang and Lu, Weiming and others},
  journal={arXiv preprint arXiv:2507.15846},
  year={2025}
}

@misc{yang2025realfactbench,
      title={RealFactBench: A Benchmark for Evaluating Large Language Models in Real-World Fact-Checking}, 
      author={Shuo Yang and Yuqin Dai and Guoqing Wang and Xinran Zheng and Jinfeng Xu and Jinze Li and Zhenzhe Ying and Weiqiang Wang and Edith C. H. Ngai},
      year={2025},
      eprint={2506.12538},
      archivePrefix={arXiv},
      primaryClass={cs.CL},
}

@misc{yang2025rama,
    title={RAMA: Retrieval-Augmented Multi-Agent Framework for Misinformation Detection in Multimodal Fact-Checking},
    author={Shuo Yang and Zijian Yu and Zhenzhe Ying and Yuqin Dai and Guoqing Wang and Jun Lan and Jinfeng Xu and Jinze Li and Edith C. H. Ngai},
    year={2025},
    eprint={2507.09174},
    archivePrefix={arXiv},
    primaryClass={cs.CL}
}

@misc{xie2025dualagentadversarialframeworkrobust,
      title={A Dual-Agent Adversarial Framework for Robust Generalization in Deep Reinforcement Learning}, 
      author={Zhengpeng Xie and Jiahang Cao and Yulong Zhang and Qiang Zhang and Renjing Xu},
      year={2025},
      eprint={2501.17384},
      archivePrefix={arXiv},
      primaryClass={cs.LG},
      url={https://arxiv.org/abs/2501.17384}, 
}

@article{li2025search,
  title={Search-o1: Agentic search-enhanced large reasoning models},
  author={Li, Xiaoxi and Dong, Guanting and Jin, Jiajie and Zhang, Yuyao and Zhou, Yujia and Zhu, Yutao and Zhang, Peitian and Dou, Zhicheng},
  journal={arXiv preprint arXiv:2501.05366},
  year={2025}
}

@article{qi2024webrl,
  title={WebRL: Training LLM Web Agents via Self-Evolving Online Curriculum Reinforcement Learning},
  author={Qi, Zehan and Liu, Xiao and Iong, Iat Long and Lai, Hanyu and Sun, Xueqiao and Zhao, Wenyi and Yang, Yu and Yang, Xinyue and Sun, Jiadai and Yao, Shuntian and others},
  journal={arXiv preprint arXiv:2411.02337},
  year={2024}
}

@article{chen2025research,
  title={Research: Learning to reason with search for llms via reinforcement learning},
  author={Chen, Mingyang and Li, Tianpeng and Sun, Haoze and Zhou, Yijie and Zhu, Chenzheng and Wang, Haofen and Pan, Jeff Z and Zhang, Wen and Chen, Huajun and Yang, Fan and others},
  journal={arXiv preprint arXiv:2503.19470},
  year={2025}
}

@article{li2025webthinker,
  title={WebThinker: Empowering Large Reasoning Models with Deep Research Capability},
  author={Li, Xiaoxi and Jin, Jiajie and Dong, Guanting and Qian, Hongjin and Zhu, Yutao and Wu, Yongkang and Wen, Ji-Rong and Dou, Zhicheng},
  journal={arXiv preprint arXiv:2504.21776},
  year={2025}
}

@article{song2025r1,
  title={R1-Searcher: Incentivizing the Search Capability in LLMs via Reinforcement Learning},
  author={Song, Huatong and Jiang, Jinhao and Min, Yingqian and Chen, Jie and Chen, Zhipeng and Zhao, Wayne Xin and Fang, Lei and Wen, Ji-Rong},
  journal={arXiv preprint arXiv:2503.05592},
  year={2025}
}

@article{jin2025search,
  title={Search-r1: Training llms to reason and leverage search engines with reinforcement learning},
  author={Jin, Bowen and Zeng, Hansi and Yue, Zhenrui and Yoon, Jinsung and Arik, Sercan and Wang, Dong and Zamani, Hamed and Han, Jiawei},
  journal={arXiv preprint arXiv:2503.09516},
  year={2025}
}

@article{deepresearcher,
  title={Deepresearcher: Scaling deep research via reinforcement learning in real-world environments},
  author={Zheng, Yuxiang and Fu, Dayuan and Hu, Xiangkun and Cai, Xiaojie and Ye, Lyumanshan and Lu, Pengrui and Liu, Pengfei},
  journal={arXiv preprint arXiv:2504.03160},
  year={2025}
}

@article{shao2024deepseekmath,
  title={Deepseekmath: Pushing the limits of mathematical reasoning in open language models},
  author={Shao, Zhihong and Wang, Peiyi and Zhu, Qihao and Xu, Runxin and Song, Junxiao and Bi, Xiao and Zhang, Haowei and Zhang, Mingchuan and Li, YK and Wu, Y and others},
  journal={arXiv preprint arXiv:2402.03300},
  year={2024}
}

@article{team2023gemini,
  title={Gemini: a family of highly capable multimodal models},
  author={Team, Gemini and Anil, Rohan and Borgeaud, Sebastian and Alayrac, Jean-Baptiste and Yu, Jiahui and Soricut, Radu and Schalkwyk, Johan and Dai, Andrew M and Hauth, Anja and Millican, Katie and others},
  journal={arXiv preprint arXiv:2312.11805},
  year={2023}
}

@article{kaelbling1996reinforcement,
  title={Reinforcement learning: A survey},
  author={Kaelbling, Leslie Pack and Littman, Michael L and Moore, Andrew W},
  journal={Journal of artificial intelligence research},
  volume={4},
  pages={237--285},
  year={1996}
}

@article{yang2025qwen25,
  title={Qwen2. 5 technical report.},
  author={Yang, An and Yang, Baosong and Zhang, Beichen and Hui, Binyuan and Zheng, Bo and Yu, Bowen and Li, Chengyuan and Liu, Dayiheng and Huang, Fei and Wei, Haoran and others},
  journal={arXiv preprint arXiv:2412.15115, 2024.},
  year={2024}
}

@article{yang2025qwen3,
  title={Qwen3 technical report},
  author={Yang, An and Li, Anfeng and Yang, Baosong and Zhang, Beichen and Hui, Binyuan and Zheng, Bo and Yu, Bowen and Gao, Chang and Huang, Chengen and Lv, Chenxu and others},
  journal={arXiv preprint arXiv:2505.09388},
  year={2025}
}

\end{document}